\documentclass{pazha2}
\usepackage{graphicx}
\usepackage{color}
\usepackage{multirow}
\definecolor{darkblue}{rgb}{0,0,0.9}

\def\*{$^{*}$}

\sloppypar

\begin{document}
\journalinfo{2017}{43}{10}{656}{727}{735}[663]

\title
{\Large IGR J17445-2747 --- YET ANOTHER X-RAY BURSTER\\ IN THE GALACTIC BULGE}

\author{I. A. Mereminskiy\address{1}\email{i.a.mereminskiy@gmail.com},
      S. A.~Grebenev\address{1},
      R. A.~Sunyaev\address{1,2}\\ [4mm]
$^1${\it Space Research Institute, Russian Academy of  Sciences, Moscow, Russia}\\
\noindent
$^2${\it Max-Planck-Institut f\"ur Astrophysik, Garching, Germany}}

\vspace{2mm}
\submitted{May 30, 2017}
\shortauthor{MEREMINSKIY et al.}
\shorttitle{IGR J17445-2747 --- YET ANOTHER X-RAY BURSTER}

\begin{abstract}
\noindent
The discovery of a type I X-ray burst from the faint
unidentified transient source \mbox{IGR\,J17445-2747} in the Galactic
bulge by the \mbox{JEM-X} telescope onboard the INTEGRAL
observatory is reported. Type I bursts are believed to be
associated with thermonuclear explosions of accreted matter on
the surface of a neutron star with a weak magnetic field in a
low-mass X-ray binary. Thus, this observation allows the nature
of this source to be established.

\noindent
{\bf DOI:} 10.1134/S106377371710005X

.
\end{abstract}


\section*{INTRODUCTION}
\noindent
The X-ray source IGR\,J17445-2747 was discovered by the
IBIS/ISGRI gamma-ray telescope (Lebrun et al. 2003; Ubertini et
al. 2003) onboard the INTEGRAL observatory (Winkler et al. 2003)
when analyzing the combined sky images (mosaics) obtained in the
first two years of its in-orbit operation.  In the second
catalog of sources detected by this telescope (Bird et al. 2006)
it is designated as a faint unidentified persistent source with
a flux of $(6.8\pm0.8) \times 10^{-12}\ \mbox{erg s}^{-1}
\mbox{cm}^{-2}$ in the 20--40 keV energy band. In the succeeding
catalogs (Bird et al. 2007, 2010; Krivonos et al. 2007, 2010)
the source is already marked as a highly variable one. It was
most active in February-March 2004 (Krivonos et al. 2010), when
its 17--60 keV flux reached $(3.9\pm0.3)\times
10^{-11}\ \mbox{erg s}^{-1} \mbox{cm}^{-2}$. The second
telescope onboard the INTEGRAL observatory, JEM-X (Lund et
al. 2013), sensitive in a softer (standard) X-ray band did not
detect this outburst (Grebenev and Mereminskiy 2015).

Since the discovery of the source, attempts have been repeatedly
made to improve its localization using focusing X-ray telescopes
with grazing-incidence mirrors and to determine its
nature. However, SWIFT-XRT (Landi et al. 2007) and
\mbox{CHANDRA} (Tomsick et al. 2008) observations did not
brought clarity: only one source, though bright but laying
outside the formal IBIS/ISGRI error circle of IGR\,J17445-2747
with a radius of 1\farcm5 at 68\% confidence, was detected
(Krivonos et al. 2007).  There was no reason to believe it to be
associated with IGR\,J17445-2747: it had already become clear
that IGR\,J17445-2747 is highly variable; therefore, it simply
could be in the off-state during these rather short
observations. Finally, in the archive of all XMM-Newton slew
observations Malizia et al. (2010) found a bright transient
source XMMSL1\,J174429.4-274609 with a peak 0.2--12 keV flux of
$\simeq1.6\times 10^{-12}\ \mbox{erg s}^{-1} \mbox{cm}^{-2}$ and
a dynamic range of variability of more than 40. Based on the
coincidence, within the error limits, of the positions of the
sources in the sky and their strong variability, Malizia et
al. (2010) suggested that XMMSL1\,J174429.4-274609 is a soft
X-ray counterpart of IGR\,J17445-2747. Unfortunately, the
peculiarities of the XMM--Newton operation in the slew mode
degrade significantly the accuracy of source localization. A
cross-correlation of the source positions from the slew catalog
with optical catalogs (Saxton et al. 2008) showed that the
90\%-confidence position error is 17\arcsec.  This makes an
optical identification of this source virtually impossible
because of the high star density near the Galactic center, where
it is located.
\begin{figure*}[t]
\centering
\includegraphics[width=0.90\textwidth]{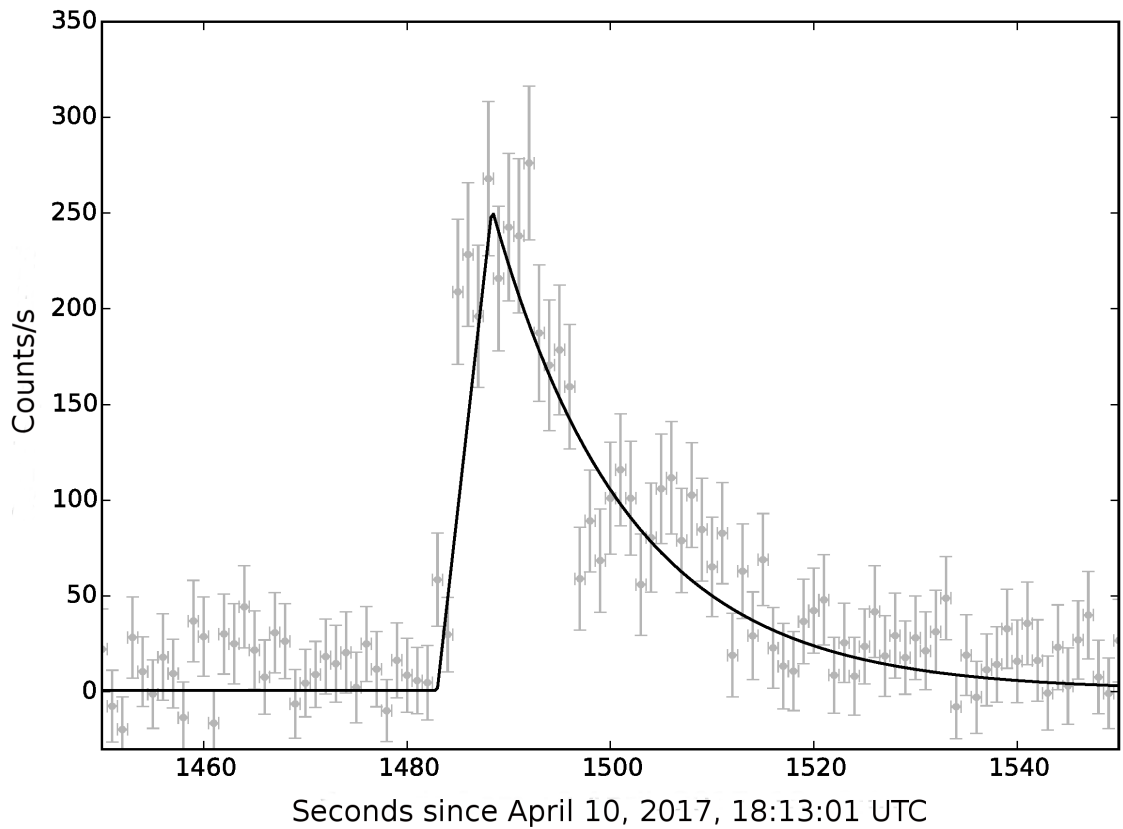}
\caption{{\rm Time profile of the X-ray burst detected from
  IGR\,J17445-2747 on April 10, 2017, based on the sum of data
  from the two \mbox{JEM-X} modules. The energy band is 3--10
  keV, the time resolution is 1~s. The solid line indicates a
  FRED (fast rise, exponential decay) model fit to the
  profile. The characteristic rise time is $\tau_r =
  5.4\ (-5.0/+0.2)$ s, the exponential decay time is $\tau_d =
  13.3\pm1.9$ s. The persistent emission from the source was
  subtracted.}}
\label{fig:detelc}
\end{figure*}

In this paper we report the detection of a type I X-ray burst
from this source, which allows it to be identified with an
accreting neutron star with a weak magnetic field in a low-mass
X-ray binary (an X-ray burster). We have already published brief
information about this event in Astronomer's Telegrams
(Mereminskiy et al. 2017a). The details of these observations
and the results of subsequent studies of this source, in
particular, its more accurate localization, are presented here.

\section*{THE BURST PROFILE AND LOCALIZATION}
\noindent
The burst was detected by the \mbox{JEM-X} telescope (or the Joint
European X-ray Monitor, Lund et al. 2003) onboard the INTEGRAL
observatory on April 10, 2017, during the scanning observations
of the Galactic center region carried out at the request of
R.A. Sunyaev (see Krivonos et al. 2012). This telescope is
sensitive in the standard 3--35 keV X-ray band. It allows the
sky to be imaged on the principle of a coded aperture in a field of
view with a diameter of 13\fdg2 FWZR (the diameter of the fully
coded region is 4\fdg8) limited by the collimator. A gas chamber
with an entrance window area of $\sim490\ \mbox{cm}^2$ and an
energy resolution $\Delta E/\,E \sim 16$\% FWHM at 6 keV is used
as a detector. The effective area at the center of the field of
view is only $\simeq75\ \mbox{cm}^2$, because more than 80\% of
the detector is blocked by the opaque mask and collimator
elements. There are two identical telescope modules onboard; if
they operate simultaneously, then the effective area is twice as
large, $\simeq 150\ \mbox{cm}^2$. The \mbox{JEM-X} telescope has a
fairly high angular resolution ($\sim3\farcm35$ FWHM), which is
important in investigating sky regions densely populated by
X-ray sources (such as the Galactic center and bulge).

Figure\,1 shows the burst time profile (the time dependence of
the photon count rate). We see that it was characterized by a
fast linear rise in count rate at the beginning and a slow
exponential decay at the end. Fitting by the corresponding model
(FRED) (the solid line in the figure) gave the characteristic
rise and decay times of the count rate $\tau_r = C_{\rm
  max}/(dC/dt) = 5.4\ (-5.0/+0.2)$ s and $\tau_d = 13.3\pm1.9$
s. The maximum occurred at 18\uh37\um49\us\ UTC. The burst was
detected by both \mbox{JEM-X} modules, and Fig.\,1 presents
their combined count rate. In Fig.\,2 the count rates of the two
modules are given separately and with a higher time resolution
($0.5$ s).
\begin{figure*}[t]
  \centering
\hspace{-0.2cm}\includegraphics[scale=0.65]{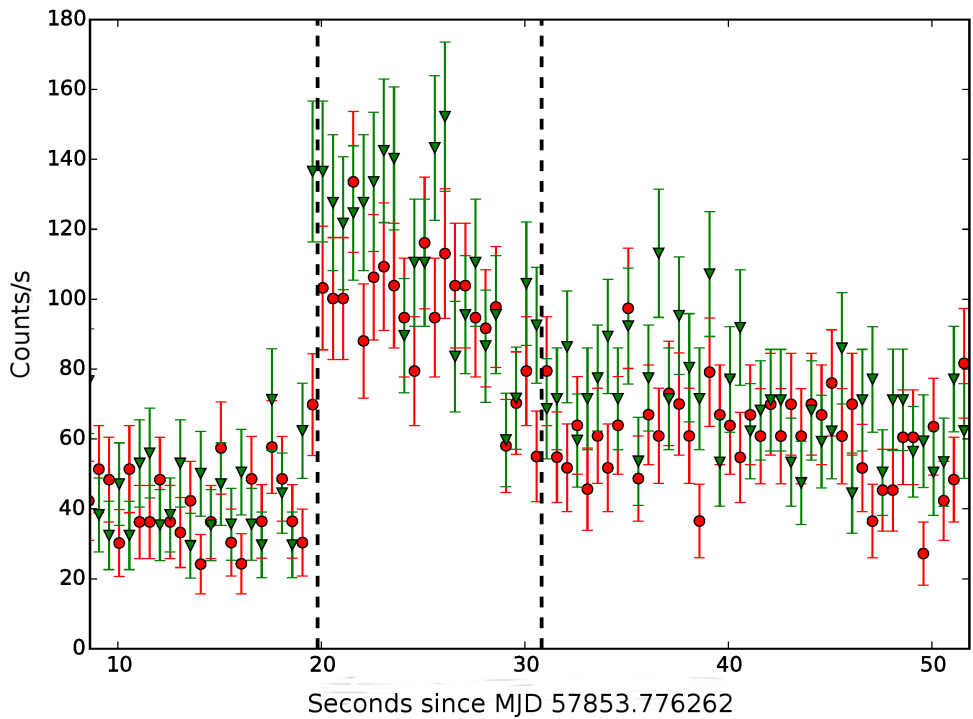}
\caption{\rm Variations in the count rate of the two \mbox{JEM-X}
  modules during the X-ray burst. The energy band is 3--10 keV,
  the time resolution is 0.5 s. The red circles and green
  triangles indicate the data from the \mbox{JEM-X1} and \mbox{JEM-X2}
  modules, respectively. The vertical dashed lines bound the
  time interval in which the image was constructed to identify
  the burst source.}
\label{fig:detelc2}
\end{figure*}

To determine the burst source, we constructed sky images in the
JEM-X field of view (S/N maps) over the entire time of the
telescope's pointing at the region under consideration (scw:
180300330010, the exposure time is 1750~s) and within 13~s near
the burst maximum (the time interval used is marked in Fig. 2 by
the vertical dashed lines). The images were obtained in the
3--20 keV band. They are presented in Fig.\,3. Although many
bright known sources, including bursters, GX\,3+1, KS\,1741-293,
A\,1742-294, A\,1743-288, and also the black hole candidate
1E\,1740.7-2942 and the X-ray transient IGR\,J17445-2747, are
seen in the first image (on the left), only one of them, namely
IGR\,J17445-2747, the only source that is also present in the
second image (on the right), can be responsible for the
burst. It was detected at the level of S/N~$\simeq5$ in the left
image (over the entire observation) and at the level of
S/N~$\simeq10$ in the right one (during the burst). At this time
the 3--10 keV photon flux from it was comparable to that from
the Crab Nebula. The measured position of the burst source in
the sky, R.A. = 266\fdg133 and Decl. = --27\fdg784 (epoch
2000.0), is consistent with the position of IGR\,J17445-2747
(its soft X-ray counterpart XMMSL1\,J174429.4-274609) to within
1\arcmin, corresponding to the typical JEM-X position error of
sources with comparable significance (having the same S/N).

Note that in the combined mosaic of images obtained by
\mbox{JEM-X} on April 7--11 (within R.A. Sunyaev's observations
and the observations caried out at the requests of J. Wilms and
E. Kuulkers) the source is detected with a higher significance
at $\mbox{S/N} = 6.4$ with a 5--10 keV flux $F_X \simeq (3.2 \pm
0.5)\times 10^{-11}\ \mbox{erg s}^{-1} \mbox{cm}^{-2}$.

\begin{figure*}[th]

\hspace{-0.3cm}\includegraphics[scale=0.65]{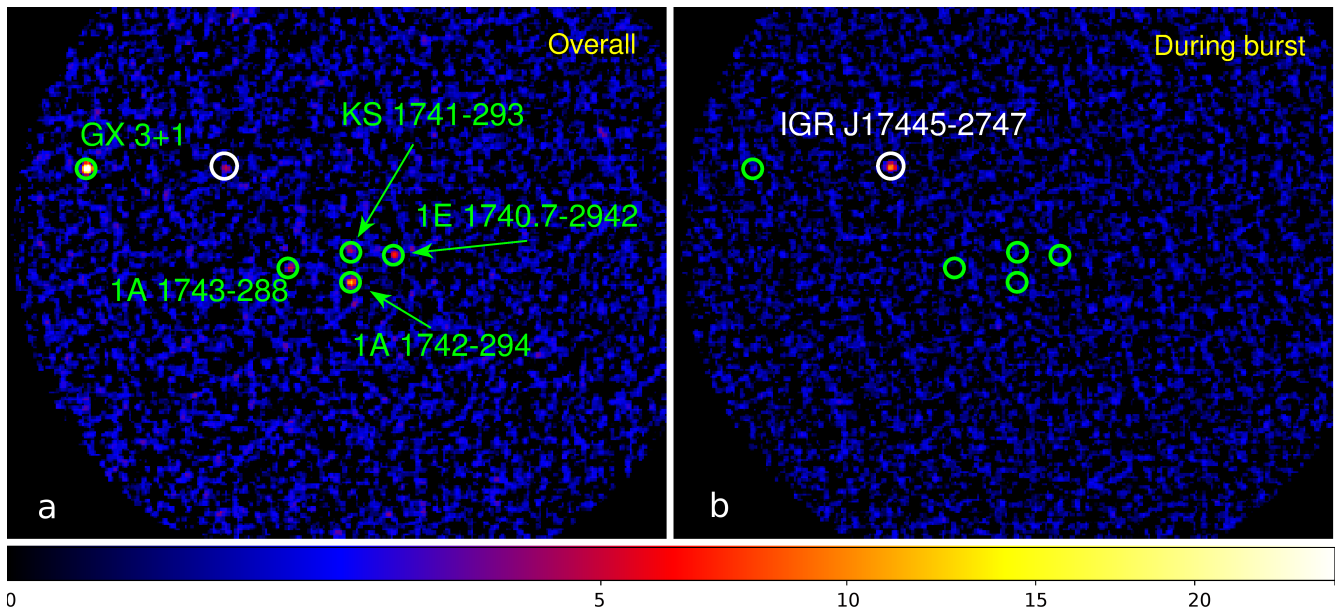}
 
 \caption{\rm JEM-X sky images (S/N maps) in the 3--20 keV band
   obtained on April 10, 2017: (a) over the entire observation
   during which the burst was detected; (b) within 13~s near the
   burst maximum. The detected bright persistent sources are
   indicated in the left image. The white circle marks
   IGR\,J17445-2747.}
\label{fig:skyconp}
\end{figure*}

\section*{IMPROVING THE POSITION BASED ON SWIFT-XRT DATA}
\noindent
Immediately after the burst detection (Mereminskiy
et al. 2017a) we initiated the SWIFT-XRT (Burrows et al. 2005) observations
of IGR\,J17445-274 to determine its accurate
position. Four observations were performed: on
April 14 (four days after the burst), May 1, 6, and 15.
The source in the field of view was detected only
during the first two observations (ID: 00035353003,
00035353005) with a mean photon count rate of
$0.80\pm0.04$ and $0.10\pm0.01\ \mbox{counts s}^{-1}$, respectively
(Mereminskiy et al. 2017b). These observations
allowed the source position to be improved
(Mereminskiy et al. 2017b). Kennea et al. (2017)
pointed out that the source position could be improved
even more by reducing the systematic errors
if the images obtained by SWIFT-UVOT simultaneously
with SWIFT-XRT are used to tie to the
known positions of bright stars (Goad et al. 2007;
Evans et al. 2009). Kennea et al. (2017) determined
the source position, R.A. = 266\fdg12647 and Decl. = 
-27\fdg76685, using only the first observation. Subsequently,
we repeated this procedure based on two
observations of the source during which it was bright
and found its coordinates: R.A. = 266\fdg12649 and
Decl. = -27\fdg76686 (epoch 2000.0, an error of 2\farcs1,
a 90\% confidence interval). Our measurements are
consistent, within the error limits, with the results of
Kennea et al. (2017).

Figure 4 shows the image (the distribution of the number of
detected photons) of the sky field around IGR\,J17445-2747
obtained from its first two SWIFT-XRT observations (on April 14
and May 1, 2017).  The image was taken in the 0.3-10 keV band. A
bright source is clearly seen at the center. The error circles
of the source localization by different telescopes are
superimposed: the orange circle is for XMM-Newton from the data
of the {\it First XMM-Newton slew catalog\/} (Saxton et
al. 2008), the yellow circle is for the same observatory from
the data of the recently published {\it Second XMM-Newton slew
  catalog}\footnote{see
  www.cosmos.esa.int/web/xmm-newton/xmmsl2-ug}, and the
green circle is from the SWIFT-XRT data obtained during the
Galactic bulge survey by this telescope (Heinke et al. 2017).

Note a significant change (and improvement) of the source
position in the {\it Second XMM-Newton slew catalog\/} compared
to the {\it First slew catalog}. It should be noted also that the
source hardness was at the level of 0.63 that is much higher
than the mean hardness of sources in the catalog (see {\sl
  www.cosmos.esa.int/documents/332006/544078/ hrplot.gif\/}). At
the same time, all the mentioned error circles capture the
source only at the very edge, implying the possibility of the
presence of other close transient sources distorting the
localization results in this region. The small blue circle
indicates the source position error from our
observations. Obviously, this localization is much better than
the previous ones and is the only one that allows an optical
identification of the source to be attempted. Such attempts have
already been made (Shaw et al. 2017; Chakrabarty et al. 2017),
but the result obtained cannot yet be deemed unambiguous.
\begin{figure*}[t]
  \centering 
\includegraphics[width=12.6cm]{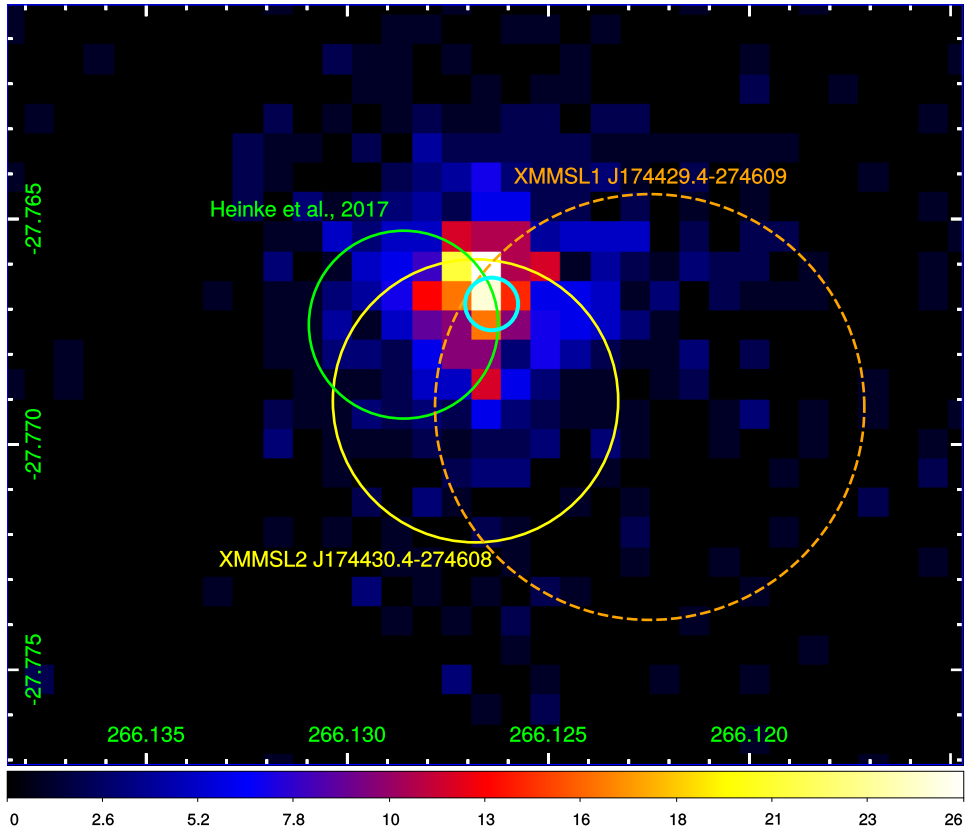}
 \caption{\rm SWIFT-XRT sky image in the 0.3--10 keV band from
   the sum of the first two observations of IGR\,J17445-2747
   (April 14 and May 1, 2017, the total exposure time is
   $\sim2100$~s). The large dashed (orange) circle, the yellow
   circle, the green circle, and the small blue circle
   represent, respectively, the error circles of the source
   localization from the data of the {\it First XMM-Newton slew
     catalog\/} (Saxton et al. 2008), the data of the {\it
     Second XMM-Newton slew catalog}, the data by Heinke et
   al. (2017), and our observational data (see also Kennea et
   al. 2017). The image is shifted from that directly measured
   by the telescope to correspond the improved coordinates.}
\label{fig:skyxrt}
\end{figure*}

Apart from a more accurate localization of the source, the
SWIFT-XRT observations allow the interstellar extinction in its
direction to be estimated.  For this purpose, we used the
radiation spectra taken in the first two observations, on April 14
and May 1. The spectra were simultaneously fitted by the {\tt
  TBabs*powerlaw} model from the XSPEC package (Arnaud et
al. 1996), the slope and normalization of the power law for each
spectrum were independent parameters, while the extinction was
assumed to be the same. The abundances were assumed to be solar
(Wilms et al. 2000), the cross-sections were taken from Verner
et al. (1996), the Cash (1979) statistic was used for the
fitting. In this way we managed to obtain a constraint on the
hydrogen column density toward the source, $N_{\rm H}\simeq
(5.6\pm 1.3)\times 10^{22}\ \mbox{cm}^{-2}$.  The slopes of the
spectra (photon indices) agree between themselves: $\alpha
\simeq2.06\pm 0.42$ and $2.16\pm0.64$ for the first and second
spectra, respectively, while the 5--10 keV fluxes were
$F_X\simeq 3.2\,(+0.2/-0.7)\times 10^{-11}$ and
$4.3\,(+0.5/-0.8)\times 10^{-12}\ \mbox{erg
  s}^{-1}\ \mbox{cm}^{-2}$, suggesting a fast decay of the
source. This is also confirmed by the absence of its detection
in the last two observations; the $3\sigma$ limit on its flux on
either of these days was $6\times 10^{-13}\ \mbox{erg
  s}^{-1}\ \mbox{cm}^{-2}.$

\section*{THE BURST SPECTRUM AND THE DISTANCE TO THE SOURCE}
\noindent
The profile of the X-ray burst from IGR\,J17445-274 has a shape
typical for type I bursts: a fast ($\la5$~s) rise and a much
slower ($\sim 15$~s) decay. Such bursts are a sign of a
thermonuclear explosion of accreted matter on the surface of a
neutron star with a weak magnetic field in a low-mass X-ray
binary. This is also confirmed by the almost blackbody spectrum
detected during the burst.
\begin{figure*}[ht]
  \centering
 \includegraphics[width=0.79\textwidth]{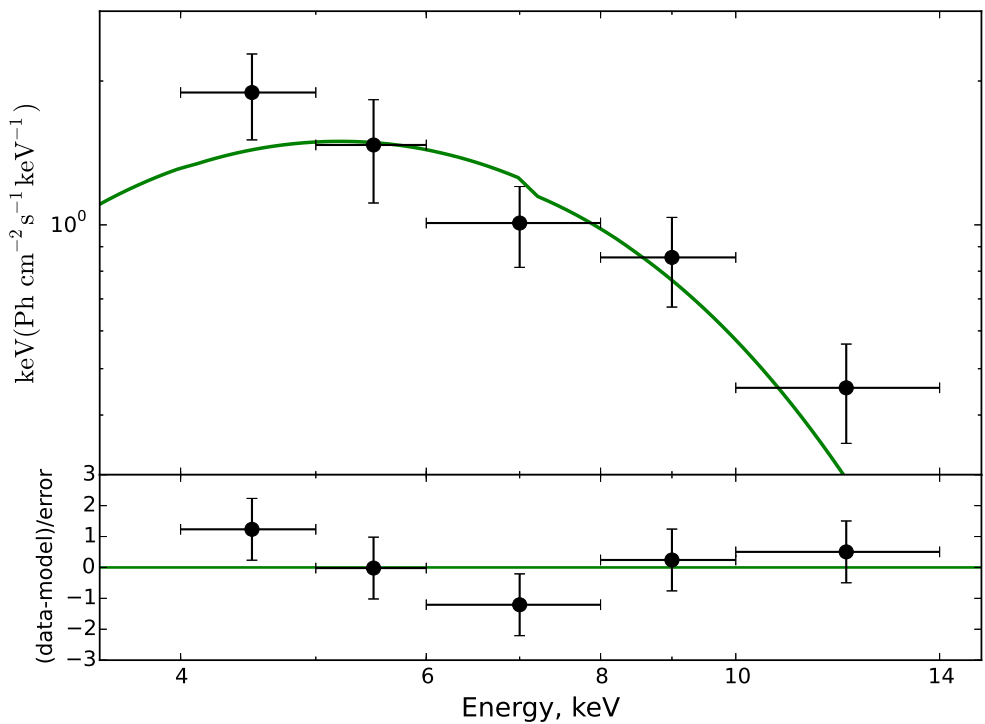}
 \caption{\rm Radiation spectrum near the burst maximum from the
   data of two \mbox{JEM-X} modules. The solid (green) line
   indicates a fit to the spectrum by a blackbody model with
   $kT_{bb}\simeq 1.6\pm0.3$ keV.}
\label{fig:spec}
\end{figure*}

We reconstructed the burst spectrum in accordance with the
procedure described in Grebenev and Mereminskiy (2015) in five
channels from 4 to 14 keV for the same time interval near
burst maximum as that used to construct the image (on the right)
in Fig.\,3 (marked by the vertical dashed lines in Fig.\,2). A
diagonal matrix based on the \mbox{JEM-X} observations of the
Crab Nebula performed within several days before the burst
detection (we selected the pointings in which the Crab Nebula
was within 3\deg\ of the center of the field of view) was used
to fit the spectrum. The resulting spectrum is presented in
Fig.\,5. We see that it is very soft, which the spectrum of a
type I burst must be.

It was fitted by the {\tt TBabs*bbodyrad} blackbody model with
absorption fixed at $N_{\rm H}$ obtained during the SWIFT-XRT
observation of the source; the blackbody temperature turned out
to be $kT_{bb}\simeq1.6\pm 0.3$ keV. Assuming that the emission
comes from the entire surface of the neutron star with a radius
$R_{bb} = 12$ km and neglecting the spectrum distortions due to
Comptonization, we can estimate the minimum distance to the
source, $d_{\rm min}\simeq5$ kpc. The total bolometric flux at
the burst maximum is $F_{bb}\simeq2.1\times10^{-8}\ \mbox{erg
  s}^{-1}\ \mbox{cm}^{-2}$.  Fitting the spectrum by the Wien
law (Grebenev et al. 2002) gives the similar temperature
$kT_W\simeq 1.60\pm 0.25$ keV and bolometric flux $F_W=L_W/(4\pi
d^2)\simeq 1.9\times 10^{-8}\ \mbox{erg s}^{-1}
\mbox{cm}^{-2}$. The photospheric radius of the neutron star in
this case occurs to be larger than the radius estimated in the
limit of a blackbody model. Under the conditions of saturated
Comptonization just the spectrum of the Wien shape must be
formed, not the Planck (blackbody) one (Kompaneets, 1957).

The limiting luminosity of bursters during bursts is known to be
restricted by the critical Eddington level on reaching which a
rapid photospheric expansion of the neutron star and an outflow
of matter begin (see, for example, Lewin et al., 1993).
Although we did not reveal any evidence of a photospheric
expansion in the burst time profile, we can estimate the maximum
distance to IGR\,J17445-2747 $d_{\rm max}\simeq 12.3$ kpc (for the
accretion of pure helium) and $d_{\rm max}\simeq 7.7$ kpc (for
the accretion of matter with solar abundances) by assuming that
the luminosity during the burst reached the Eddington level and
that the neutron star mass is equal to the standard value $M =
1.4 M_{\odot}$.

\section*{CONCLUSIONS}
\noindent
We reported the detection of a type I X-ray burst from
the poorly studied transient X-ray source IGR\,J17445-2747 in
the Galactic bulge, which allows it to be identified as a
low-mass X-ray binary where the compact object is a neutron
star. The high absorption on the line of sight, the closeness to
the Galactic center direction, and the derived lower (5 kpc) and
upper (12.3 kpc) limits on its distance make the location of
this source in the Galactic bulge most probable.

This is not the first X-ray burster discovered by the
INTEGRAL observatory. Previously, two hitherto unknown bursters,
AX\,J1754.2-2754 and IGR\,J17380-3749, were revealed/discovered
in the IBIS/ISGRI archival data by Chelovekov and Grebenev
(2007, 2010) within the framework of a special program aimed at
searching for type I X-ray bursts (Chelovekov et al. 2006, 2017;
Chelovekov and Grebenev 2011). Five more bursters,
XTE\,J1739-285, IGR\,J17254-3257 (1RXS\,J172525.5-325717),
IGR\,J17464-2811 IGR\,J17597-2201, and 1RXS\,J180408.9-342058,
were found by Brandt et al. (2005, 2006a, 2006b, 2007) and
Chenevez et al. (2012) in the \mbox{JEM-X} data. In the case of
IGR\,J17380-3749 and IGR\,J17464-2811 new sources were
discovered due to the bursts, in the other cases the origin of
the already known but unidentified sources was recognized
(IGR\,J17597-2201 was discovered by the same INTEGRAL
observatory but a few years before the burst observation, see
Lutovinov et al. 2003). These discoveries show that many as yet
unknown bursters with a low accretion (and luminosity) level can
hide in the Galaxy. Years must elapse for the critical mass of
matter needed for a thermonuclear explosion to be accumulated on
their surface. X-ray bursts in these sources occur very rarely,
but this is a real chance to detect them and to identify their
nature and it should not be missed.

\section*{ACKNOWLEDGMENTS}
\noindent
This work is based on the INTEGRAL data retrieved via its
Russian and Europian science data centers and the SWIFT data
retrieved via the NASA HEASARC. We are grateful to the SWIFT
team for rapidly planning and performing the observations in
response to our request, and to R.A. Krivonos and
I.V. Chelovekov for useful discussions. The study was
financially supported by the ``Transitional and Explosive
Processes in Astrophysics'' Subprogram of the Basic Research
Program P-7 of the Presidium of the Russian Academy of Sciences,
the Program of the President of the Russian Federation for
support of Leading Scientific Schools (grant NSh-10222.2016.2)
and the ``Universe'' theme of the scientific research program of
the Space Research Institute of the Russian Academy of Sciences.


\vspace{1cm}

\begin{flushright}
{\it  Translated by V. Astakhov\/}
\end{flushright}
\end{document}